# Supernova 1572 and other newly observed stars in the literature of the time

Amelia Carolina Sparavigna
Politecnico di Torino

SN 1572, also known as the Tycho's Nova, was a supernova largely reported and discussed in the literature of the time. Here we talk about this literature. In the latest texts, we find also mentioned the Kepler's Nova, today known as SN 1604, and other newly observed stars. We discuss them too.

**Introduction**
SN 1572, also known as the Tycho's Nova, was a supernova that appeared in the constellation Cassiopeia in November 1572. It was important because visible to the naked eye (actually the invention of the optical telescope happened during the first decade of the 1600s). The appearance of the supernova was reported in historical records and received a great attention from astronomers. The "stella nova", this was at the time the name of such a star, helped to change the models of the heavens and to speed on the revolution of astronomy, challenging the Aristotelian dogma of the unalterable nature of the heaven of the stars [1]. In the literature of the time, it is told that the star was visible from 8 November, according to the Julian Calendar (in the Gregorian Calendar, it was November 18). Tycho Brahe saw it on November 11. The supernova remained visible to the naked eye into 1574, gradually fading until it disappeared from view.
The report of the appearance of the nova started the hunt for new stars, that is, of stars "never before seen in the life or memory of anyone". Some of the newly observed stars were probably variable stars. After thirty-three years from Tycho's Nova, another supernova appeared. It was SN 1604, also known as the Kepler's Nova. It appeared in the constellation Ophiuchus, and it was unquestionably observed by the naked eye [2]. The Kepler's Star was so bright that it was visible during the day for over three weeks.
Here we discuss some literature about SN 1572, written and published in the years following its appearance, by astronomers and scholars of the time. In the latest texts, we find also mentioned the Kepler's Nova, such some other new stars. We discuss them too.

**SN 1572**
Tycho Brahe wrote an extensive work *De nova et nullius aevi memoria prius visa stella* (Concerning the Star, new and never before seen in the life or memory of anyone), that was published in 1573 [3]. Tycho tells us in his book that *Anno praecedente – that is 1572 –, mense Novembrj, die eiusdem undicesimo, vesperi post Solis occasum, cum meo more sidera coelo sereno contemplarer, novam quandam & inusitatam, praeque aliis admodum conspicuam, iuxta capitis verticem, animadverti fulgere Stellam: cumque mihi, qui inde ferè à pueritia, omnia coelj sidera perfectè (non enim magna huic scientiae inest difficultas) cognita haberem, satis evidenter constaret, nullam in eo coelj loco unquam antea extitisse, vel minimam, nedum tam conspicuae claritatis stellam: in tantam rei istius admirationem sum adductus, ut de fide, propijs oculis adhibenda, dubitare non puduerit.* The incipit of the book is remarkable and clearly explain the surprise of the young astronomer in seeing a new star.
The Figure 1 shows the map accompanying the text. This image is in the chapter *de huius novae stellae in coelo, quo ad fixas positu, et ipsius quo ad Zodiacum, longitudine et latitudine*. Tycho tells that *Conspiciebatur haec recens nata stella in Borealj coelj plaga, versus polum Arcticum, iuxta constellationem, quam veteres Magi Cassiopeam appellarunt, vicina parvae istj stellae, quae est in Cathedrae medio loco, modicum ab ea versus Cepheum remota. Constituebat etiam cum suprema Cathedrae, et ea quae in pectore Schedir appellatur, eaque quae iuxta*

*incurvationem ad ilia tendit, figuram quadrilateram. Sed ut tota res melius cognoscatur, praecipuas stellas sideris Cassiopeae una cum huius novae ad illas positu, oculis subijciam* [3]. As explained in [4], a lack of detectable parallax forced the astronomer to locate the stella nova beyond the sphere of the Moon and of the planets, that is, in the celestial realm, supposedly unalterable according to Aristotelian doctrine. The publication of this discovery made Brahe a well-known name among scientists across Europe.

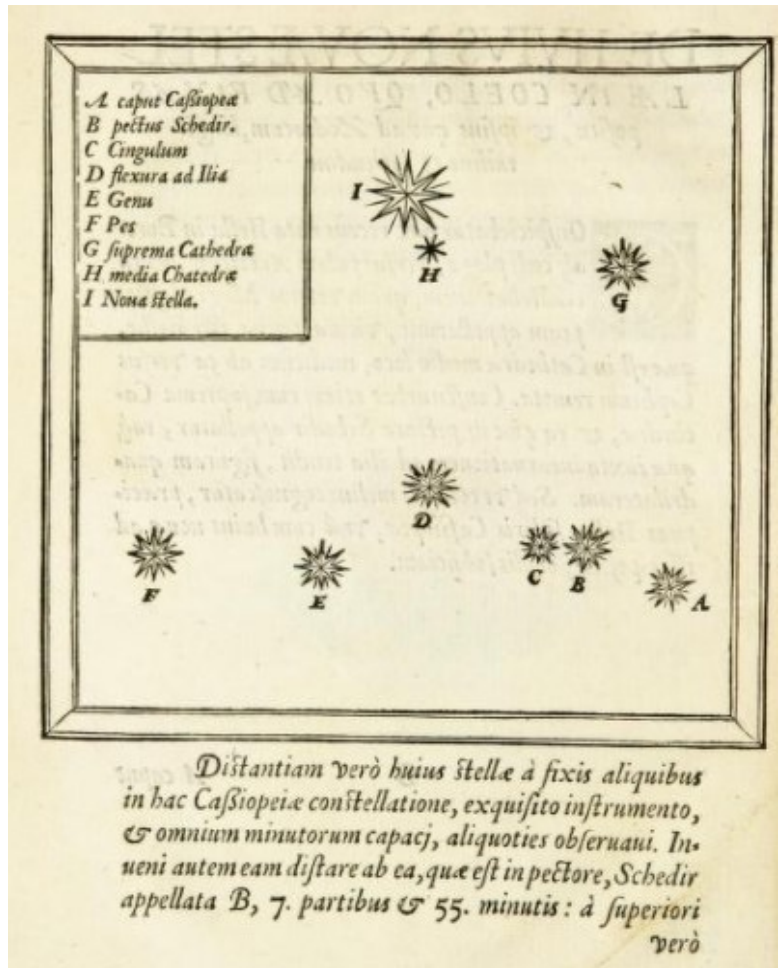
Figure 1: The supernova in Tycho's map.

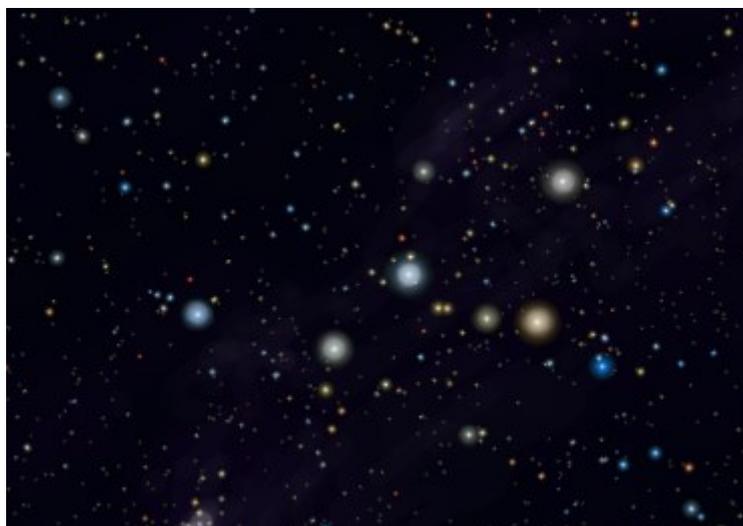
Figure 2: Here an image of Cassiopeia from a picture by Roberto Mura, 2009, for Wikipedia.

If we compare Figure 1, which is giving the chart depicted by Brahe, and Figure 2, a modern image of Cassiopeia, we can easily appreciate the accuracy of the astronomer in preparing it. Actually, "there are many allusions in the literature to Tycho's great advances in the accuracy of astronomical observations; Kepler's unhesitating belief in Tycho's accuracy is only the most famous example" [5]. Let us add that the astronomer made the observation of the supernovae when he was 26 years old.

As told in [1], "Tycho was not the first to observe the 1572 supernova, although he was probably the most accurate observer of the object. Almost as accurate were his European colleagues, such as Wolfgang Schuler, Thomas Digges, John Dee, Francesco Maurolico, Jerónimo Muñoz, Tadeáš Hájek, or Bartholomäus Reisacher".

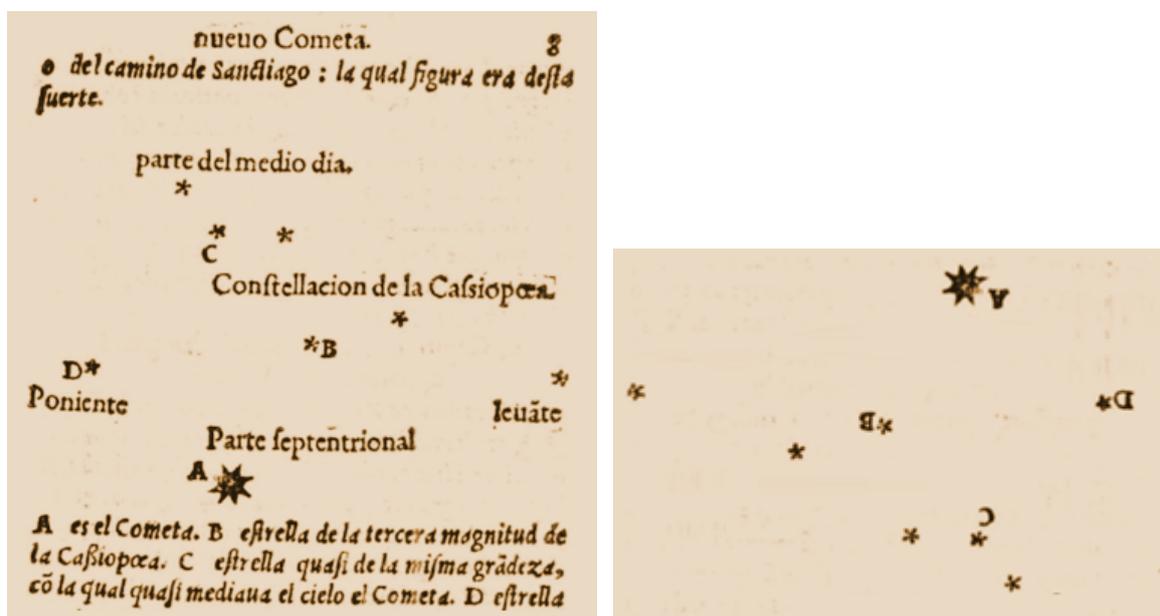

Figure 3: The map of the stars of constellation Cassiopeia in the book by Jerónimo Muñoz on the left. On the right, the same map rotated for comparison with Figures 1 and 2.

Thanks to GREDOS, Repositorio documental Universidad de Salamanca, at the link gredos.usal.es/jspui/handle/ 10366/83106, we can see the text of Jerónimo Muñoz [6]. In the Figure 3, we see a part of the page, showing the position of the "Cometa" in Cassiopeia. Muñoz tells us that *Soy cierto que el segundo día de noviembre, 1572, no havía este cometa en el cielo, porque, de propósito, más de hora y media después de las seys de la tarde, enseñé en Hontiñente a muchas personas públicamente a conocer las estrellas. Y havía pastores muy exercitados en ellas, los quales me avisaron a los 18 que por la mañana aparecía una nueva estrella. A los dos de deziembre, de propósito mirando el cielo, vide cerca de la Cassiopea una estrella como el Luzero; ... me pareció que este cometa començó a hazerse a los 31 días, 22 horas de noviembre. Examiné esto por relaciones de calcineros y pastores que están sobre Torrente, y averigüé que, a 11 o a 12 de noviembre, la començaron a ver. La magnitud aparente d'él parecía entonces algo mayor que la de Júpiter, que distava del cometa 59 grados, y casi ygualava con la del Luzero que por la mañana aparecía.*

Muñoz is giving the position of the nova by a map (Figure 3). If we compare this map with the image in the Figure 2, and with the Brahe's chart, it is easy to find that the Brahe's map is more accurate.

At the time the new star appeared, there were several reports about it. As previously told, in England Thomas Digges observed the star [7]. Here in the Figure 4 the two pages containing the new phenomenon in Digges' book.

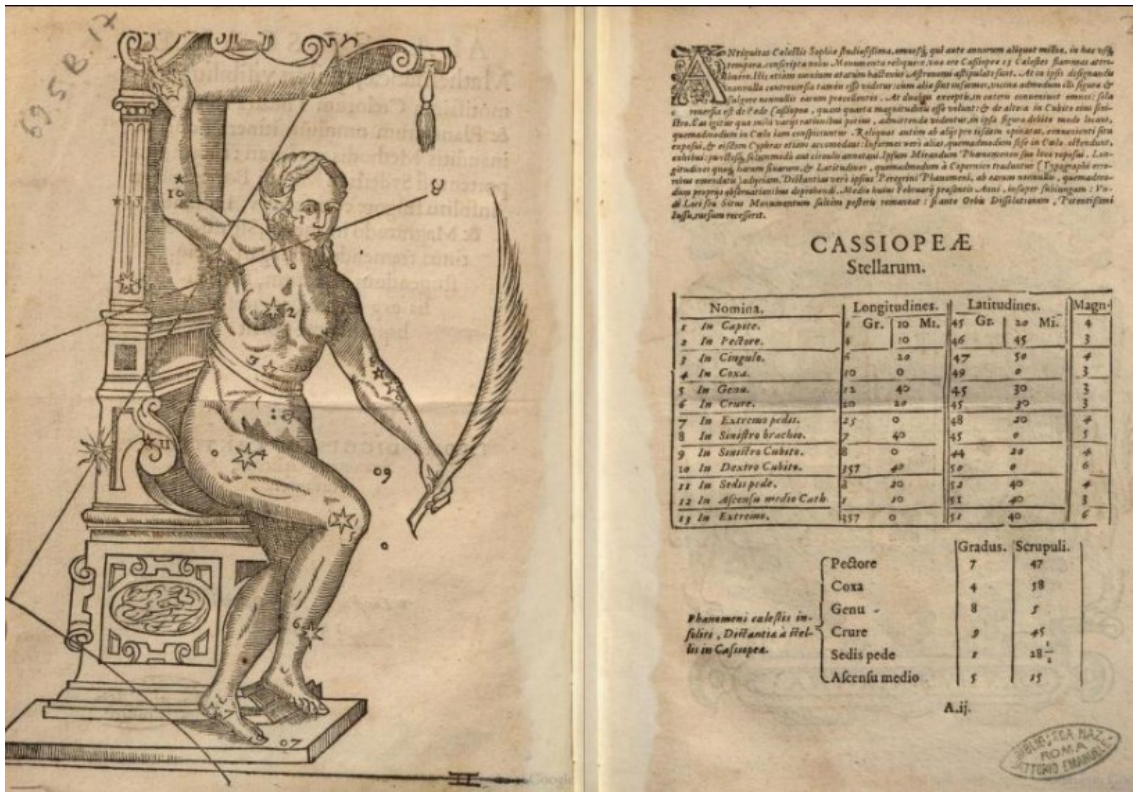

Figure 4: The first two pages of book [7], containing the description of the new phenomenon in the constellation Cassiopeia.

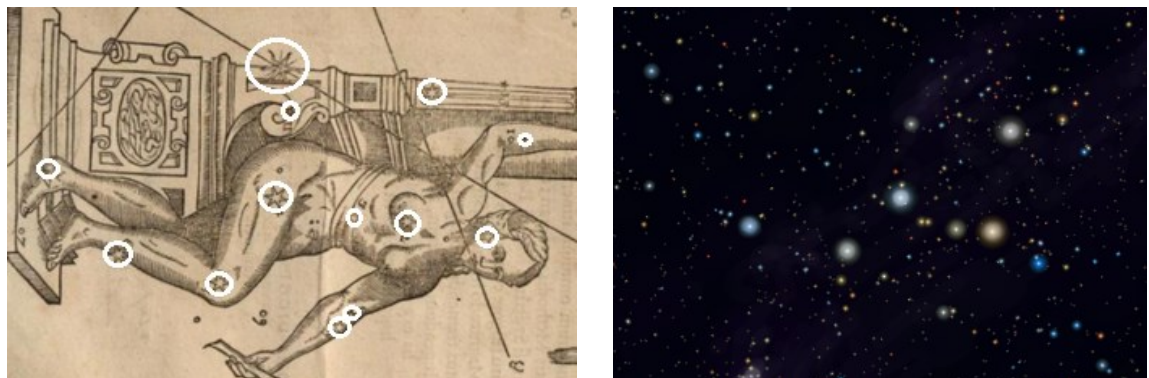

Figure 5: Here again, we can compare Cassiopeia given the Figure 4 and in the Figure 2.

We can also compare Digges' map and the modern image, as in the Figure 5. The quality of Digges' map is excellent. As explained in [8], "observations of the period, including those by Digges, have been used recently to discover the remnant of what is sometimes called 'Tycho's supernova'. Of course, Digges and his contemporaries had no concept of supernovae. For them, the 'phenomenon' (the non-committal term used by some baffles observers) was a puzzling anomaly".

The stella nova attracted the attention and observations of "countless astronomers and astrologers", of the natural philosophers and also of powerful theologians and politicians [8]. Many of them wrote and published printed and manuscript works, and therefore we could tell that the star was an editorial success.

Among the first publications on it, we find that of December 1572, in Paris, *La Declaration d'un Comete ou Estoille Prodigieuse* [8]. The author revealed his identity as L.G.D.V., that is, Jean Gosselin de Vire. The map of stars given in the Declaration (and reported in the Figure 3 of [8]) is the same that we can find in a book by Gosselin published in 1577. Here the map is given in the Figure 6 [9]. The Declaration was discussed in a Letter of 1573, imprinted in London by Thomas Marche printer, the author of which is anonymous. According to [8], the author of the Letter was Digges himself.

In the book of Jean Gosselin published in 1577 we read the follow: *Nostris autem temporibus, stellam novam vidimus maiorem stellis primae magnitudinis, & coelo Lune altiorem: quae perpetuò stetit prope stellam quartae magnitudinis quae est in dorso & in cingulo Cassiopeae. Ab hac autem stella parum distabat nova versus humeros Cephei vergens. Illam verò novam stellam observavimus, à die decimosexto Novembris anni millesimi quingentesimi septuagesimisecundi: usque in diem decimumoctavum Februarij, anni millesini quindentesimi septuagesimi quarti: quo die, Henricus Rex Poloniae, Cracoviam ingressus est. A quo die amplius non apparvit nobis ea nova stella maximè apogea facta quae inter Cepheum & Cassiopeam ita erat sita, ut hoc diagramma sequens commonstrat* [9].

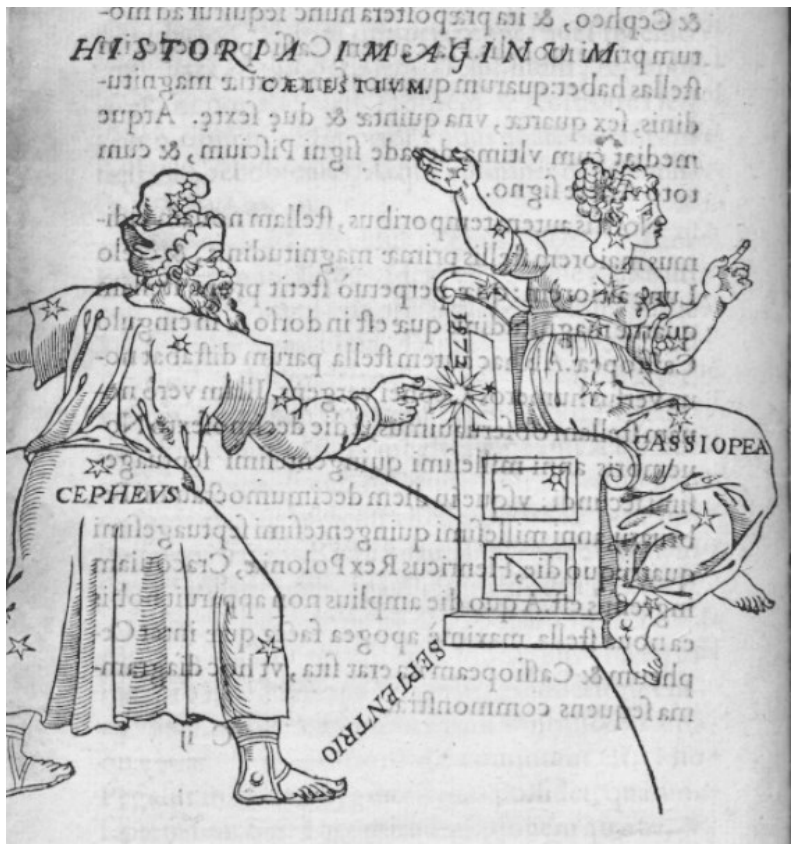

Figure 6. The position of the supernova in Gosselin's book [9].

**Searching in Google Books**

Let us consider a search "stella nova" or "stella noua" or "stella peregrina" in Google Books. Besides many copies of Brahe's book, we find several results. Here in the following some of them.

1573 - Stella Amiralia Nova Et Prodigiosa Carmine Descripta a M. Johanne Pomario Iuniore Magdeburgensi. (Johannes der Jüngere Baumgarten). It is a poem. Here the incipit:
*Nox erat, & varia sculptis in imagine stellis / Tota colorati radiabat regia coeli. / Hinc ego vel fato, vel noctis amore serenae, / Prouocor, intuitu specularer ut astra fideli. / Erectos igitur tollenti ad sidera vultus, / Ante alia offertur nullo fax cognita saeclo, / Fax nova, fax rutilans, quae sese imiiscuit astris / Cassiopeia tuis, geminisque reposta sub ursis, / Luce micat tremula, gressuque immobilis omni / Uno fixa loco, positum tutatur eundem.*

1573 - De Nova Stella. Ivdicivm Cypriani Leovitii à Leonicia, Mathematici, de noua Stella siue Cometa, viso mense Nouembri ac Decembri, Anni Domini 1572. Item mense Ianuario & Februario, Anni Domini 1573, by Cyprian Leowitz (Czech: Cyprián Karásek Lvovický, Latin: Cyprianus Leovitius) (1514? – 1574), a Bohemian astronomer, mathematician and astrologer.
*ILLUXIT NV per Stella nova, splendida ac magna, maior ullo Planeta, cuius costitutionem in Coelo die 25. Novembris, et aliquot sequnetibus diebus per instrumenti examinare coepi. Fuit diversi coloris: sub finem Novembris apparuit colore flavo atque albicante: sub initium Decembris habuit rutilum colore, atque quasi sanguineum. Postea circa mediu Decembris mixturam quandam admisit, ut de singularis coloribus illis parteciparet.*

1573 - Consideratio nouae Stellae, Qvae Mense Novembri, Anno Salvtis M.D.LXXII. In Signo Cassiopeae populis Septentrionalibus longè latéque apparuit: Scripta ad Illvstrissimvm Principem Ac D.D. Fridericum, Comitem V Virtembergensem … a Nicodemo Frischlino. Philipp Nicodemus Frischlin (1547 – 1590) was a German philologist, poet, playwright, mathematician, and astronomer, born at Erzingen, today part of Balingen in Württemberg. *Quae nova sidereo profulget in aethere stella? / Quam nitium toto spargit in orbe iubar?*

1573 - Discorso Del S. Cl. Cornelio Frangipani Sopra la Stella, che è apparsa nell'anno MDLXXII in Tramontana: Doue discorrendosi di che ella sia composta, si dicchiara grandissimi effetti, Claudio Cornelio Frangipane. *Il giorno quando mi avvidi dello esser della stella di Cassiopea, la qual sin hora da infiniti è stata scorta, il 19 di Novembre, che io mi ritrovava in Padua, dove da niuno a quel tempo se ne ragionava ….* Inside the book, we find also a poem by Francesco Molino, which ends *E poi che il ciel de tenebre fu involto, / Vidi una stella di si alto splendore, / che havea sembianza al fiammeggiar del sole.*

1573 - Dichiaratione di alcuni componimenti … Giuliano Goselini, Per P.G. Pontio. *Noua Stella amorosa a noi splendete.*

1573 - De peregrina stella quae superiore anno primum apparere coepit ... Ex Philosophiae naturalis, mysticaeque Theologiae penetralibus deprompta Iudicia. Cornelis Gemma Frisius. As told in Wikipedia, Cornelius Gemma (1535 – 1578) was a physician, astronomer and astrologer, and the oldest son of cartographer and instrument-maker Gemma Frisius. *Novus hic phosphorus, (liceat verò sie propter apparentem similitudinem appellare) coepit fulgere primu anno hoc a Christo nato 1572 nona Novembris, die domini vesperi, cum tamen observantibus proximum coeli locum die octavo, etiam sereno aethere non apparuerit.* A chart from Gemma's book in Fig. 7. "As an astronomer, Gemma is significant for his observations of a lunar eclipse in 1569 and of the 1572 supernova appearing in Cassiopeia, which he recorded on 9 November, two days before

Tycho Brahe, calling it a "New Venus." With Brahe, he was one of the few astronomers to identify the Great Comet of 1577 as superlunary" [10].

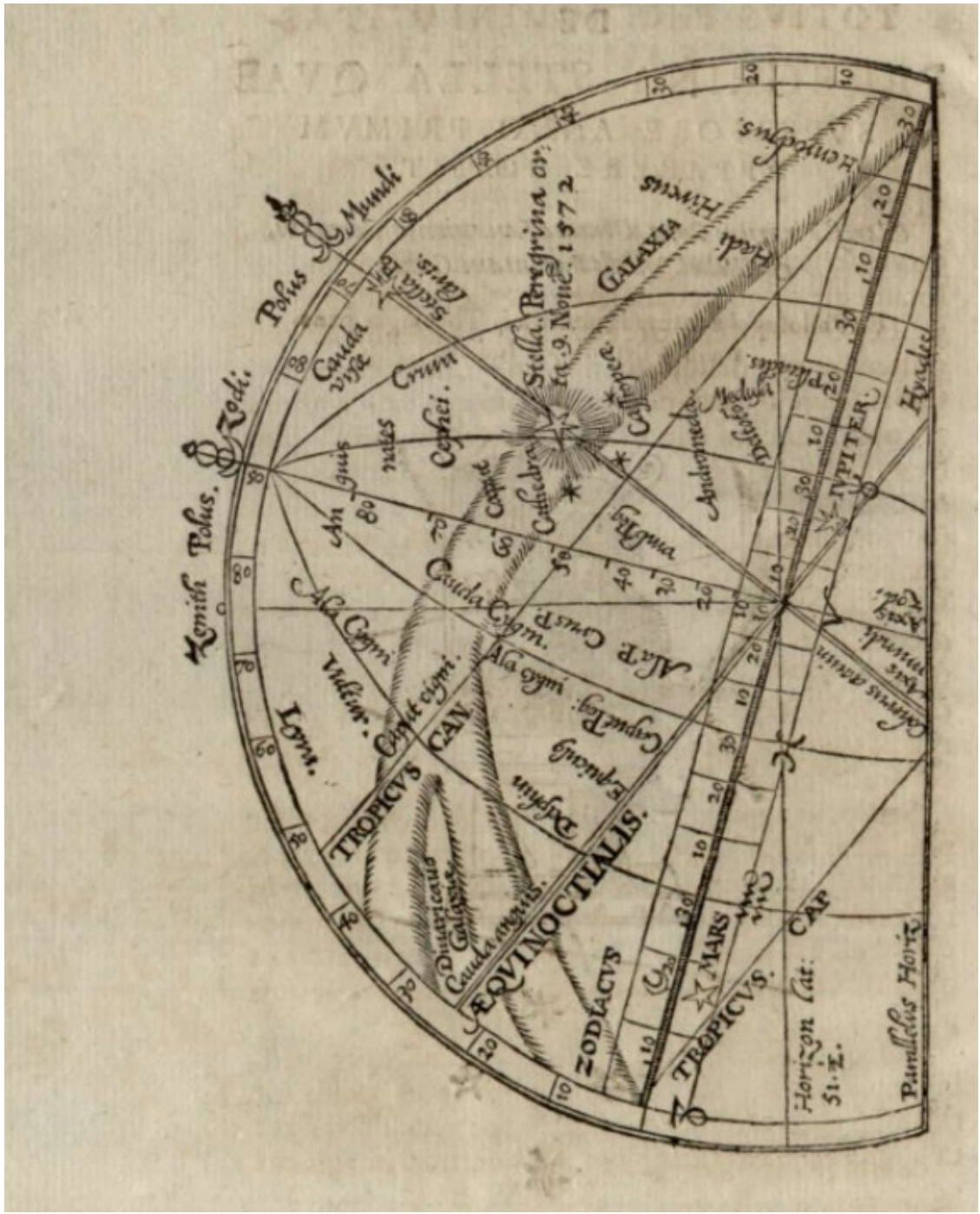

Figure 7: SN 1572 in Gemma's book.

1574 - Dialexis de novae et prius incognitae stellae: inusitatae magnitudinis & splendidissimi luminis apparitione, & de eiusdem stellae vero loco constituendo … Tadeáš Hájek z Hájku. also Also known as Thaddaeus Hagecius or Thaddeus Nemicus, Tadeáš Hájek z Hájku (1525 - 1600), was a Czech naturalist, personal physician of the Holy Roman Emperor Rudolph II and astronomer.

1576 - Responsio ad Hannibalis Raymundi scriptum: quo iterum confirmare nititur, stellam, quae anno 72 et 73. supra sesquimillesimum fulsit, non novam sed veterem fuisse, Thaddaeus Hajek ab Hayck.

1577 - De Stella inusitata et nova quae mense novembri anno 1572 conspici coepit et de Comato Sidere quod hoc mense Novembri Anno 1577 videmus a Davide Chytraeo. David Chytraeus or Chyträus (1530 - 1600) was a German Lutheran theologian and historian. He was a disciple of Melancthon. He tells us that he is talking *De Stella Nova quae die VIII Novembris Anno MDLXXII primum a nobis conspecta, totum annum prope Cassiopeam immota constitit, donec initio anni 1574 iterum evanuit.* And also, *VI. Idus Novembris primum a nobis observatum, prope Cassiopeam fulgere immotum.* (In the Introduction, we have used this date for the first observation of SN 1572).

1577 - Laurentii Gambarae Brixiani Rerum sacrarum libri tres. Idylliorum liber. vnus. Editio copiosor. Lorenzo Gambara apud haeredes Antonij Bladij, impressores camerales. *Stella quibus monstravit iter non cognita coelo Effulgens.*

1578 - Theoria Nova Coelestivm Meteōrōn, In Qva Ex Plvrivm Cometarvm Phoenomenis Epilogisticōs quædam afferuntur, de novis tertiæ cuiusdam Miraculorum Sphæræ Circulis, Polis & Axi, by Helisaeus Röslin. Roeslin (1545 – 1616) was a German physician and astrologer [11]. About the Great Comet of 1577, he concluded that it was located beyond the moon.

1579 - Catalogus nunquam antea visus, omnium cometarum secundum seriem annorum a diluvio conspectorum usque ad 1578, cum portentis seu eventuum annotationibus et de cometarum in singulis Zodiaci signis effectibus etc. Georg Caesius (a Lutheran pastor in Ansbach). *Anno 1572. Sidus quoddam iubare splendido rutilantique, in imagine Cassiopeae sub initu Novemb. vel ut alij, circa medium Octoberis, fulgere coepit, nec desunt, qui statim post interitum Admiralij Parisijs id apparvisse dicant, in fine Augusti. Duravit autem, o summu miraculu, 14. menses et ultra, ac coelo affixum semper unum et eundem locum et situm servavit.*
Here we have to remember what is told in Muñoz' book [6]. *Soy cierto que el segundo día de noviembre, 1572, no havía este cometa en el cielo.*

1579 - Poematum praeter sacra omnium libri XVII. Stephanus Myliander. *Tu noua Stella micas istis gratissima terris.*

**SN 1572 in Shakespeare's Hamlet**
*When yond same star that's westward from the pole. Had made his course to illume that part of heaven. Where now it burns, Marcellus and myself, The bell then beating one, …* (Act 1, Scene 1, Hamlet, William Shakespeare).
In [12], we can read that in 1998 researchers from Southwest Texas State University told that the star mentioned in Act 1, Scene 1, of Hamlet was SN 1572. This star was almost certainly seen by Shakespeare, then an 8-year-old schoolboy. "Dialogue between soldiers in the first scene of the play - believed to have been written in 1600 - describes a bright star and its position in the sky on a bitterly cold night. Using climate and time references in the play, the trio of researchers determined that the scene at Denmark's Elsinore Castle took place in November" [12]. Researchers are Don Olson and Russell Doescher of the Physics Department and Marilynn Olson of the English Department.
To test this fact, let us use software Stellarium that we have previously used for several simulations (see for instance, [13-17]). In the Figure 8, we see the result obtained by using

Stellarium; we see the sky above the Elsinore Castle on November 12, 1572, at 1 a.m. In azimuthal coordinates, constellation Cassiopeia is towards the west, if we are observing the sky due North. It is precisely the position of the "yond same star that's westward from the pole".

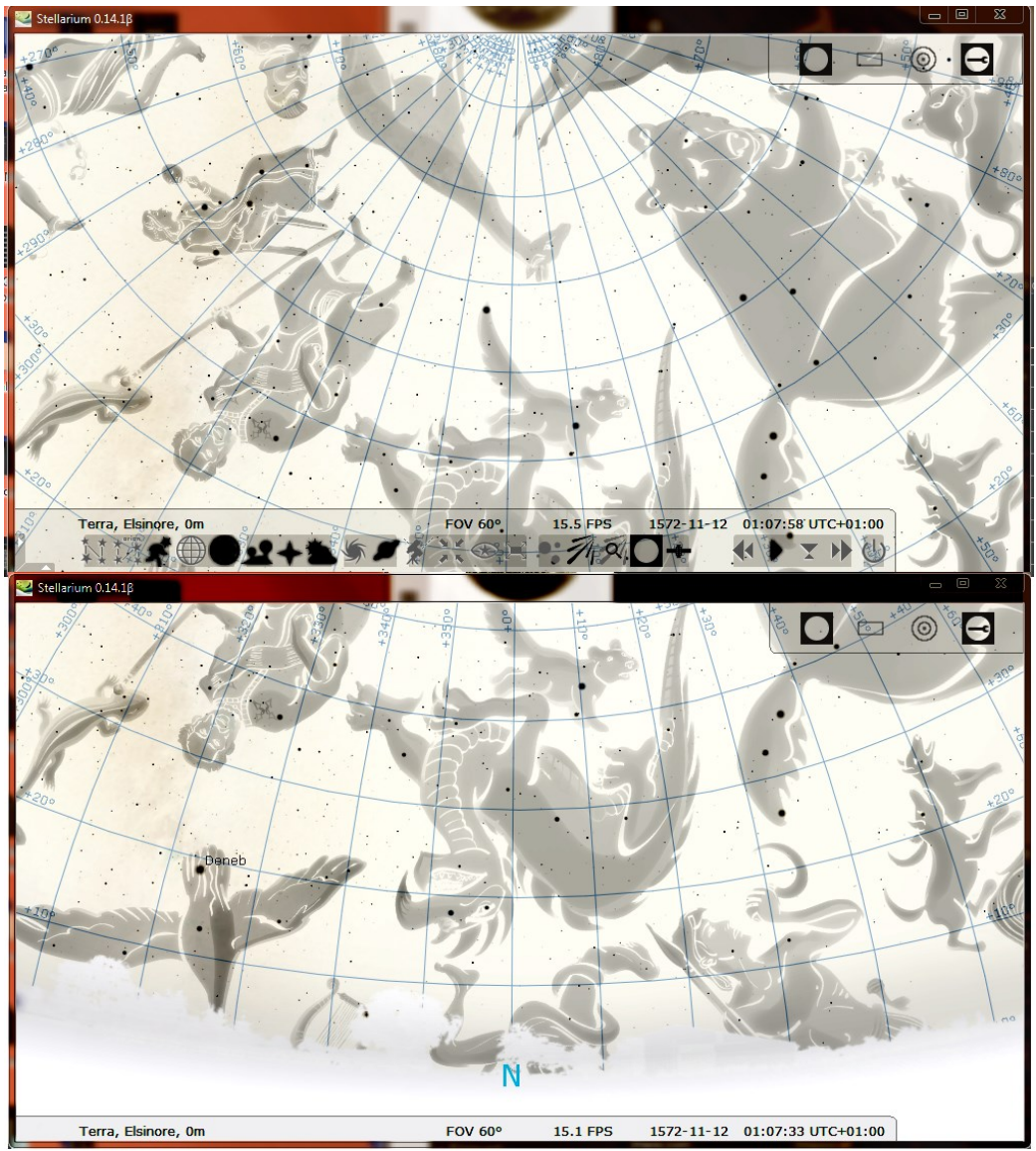

Figure 8: Simulation obtained by using Stellarium of the sky above the Elsinore Castle on November 12, 1572, at 1 a.m. The sky is represented in the azimuthal coordinates. Constellation Cassiopeia is towards the west, if we are observing the sky due North, precisely as it was the "yond same star that's westward from the pole".

**In the City of the Sun**
The City of the Sun (La Città del Sole) is a philosophical work written by Tommaso Campanella, a Dominican philosopher. The first version of the work was written in Italian in 1602, shortly after Campanella's imprisonment for heresy and sedition. In this philosophical work, we find the following: *Ma entrando l'asside di Saturno in Capricorno, e di Mercurio in Sagittario, e di Marte in Vergine, e le congiunzioni magne tornando alla triplicità prima dopo l'apparizion della stella nova in Cassiopea, sarà grande monarchia nova, e di leggi riforma e d'arti, e profeti e rinovazione. E dicono che a' cristiani questo apporterà grand'utile; ma prima si svelle e monda,*

*poi s'edifica e pianta*. As observed in [18], this is a clear reference to the supernova of 1572 in the constellation Cassiopeia, "upon which Tycho Brahe wrote De nova stella in the following year. He wrote of it again in his Astronomiae instaurata progymnasmata (published by Kepler in 1602)" [19]. "According to Firpo [Luigi Firpo] this work did not become known to Campanella until February 1611" [18]. Actually, the reference to the supernova is lacking in the earliest manuscripts.

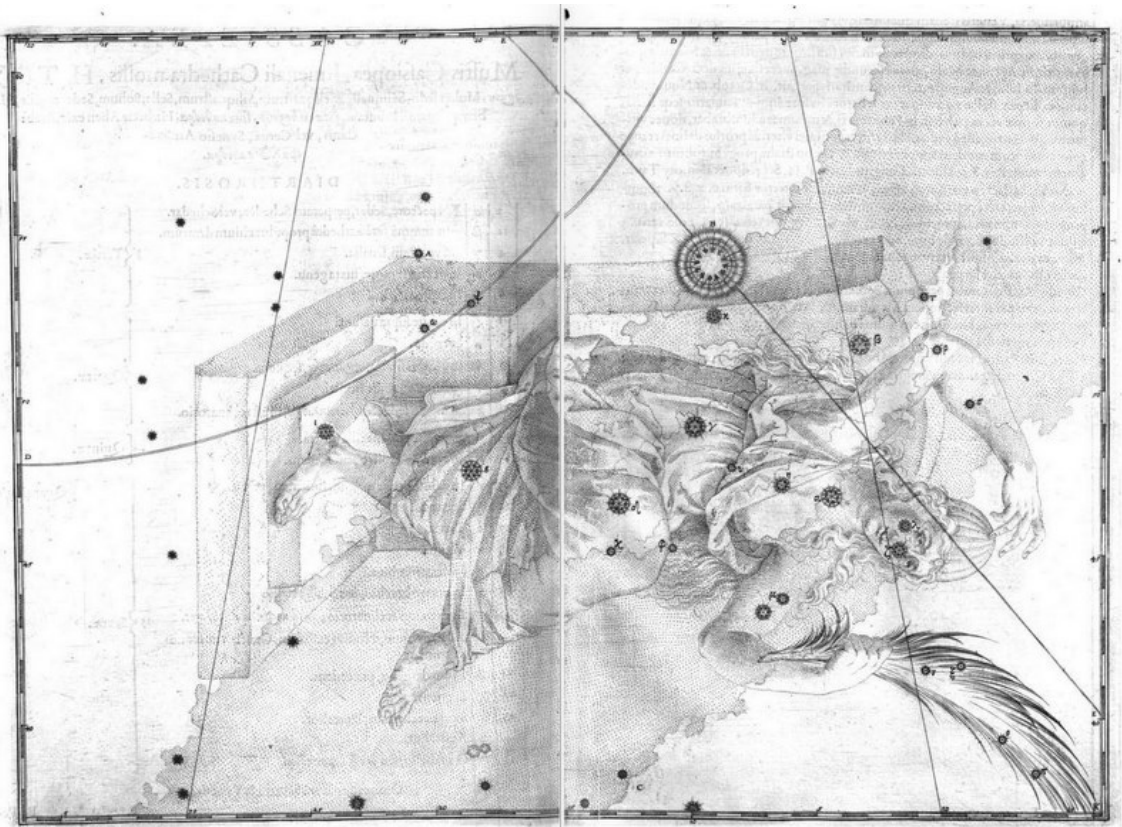
Figure 9: Cassiopeia in the Ioannis Bayeri Vranometria of 1603.

**SN 1572 in the Uranometria**
Mentioned in poetical and philosophical works, the supernova is also appearing in the charts of the stars, even after it disappeared from the sky. We can see the star in the Uranometria of 1603 by Johannes Bayer, (1572 – 1625), German lawyer and uranographer (celestial cartographer), in the constellation Cassiopeia (Figure 9). Let us note that the supernova appeared the same year that Bayer was born. Therefore, we have not to be surprised that he included it in the constellation. In [20], Bayer tells that *Peregrina Anno post Christum natum, MDLXXII, primum a.d. V. Id. Nov. animadversa, Veneris stellam quantitative visibili aequabat. Per Decembrem Iovis sidus ferme aemulabatur. Ianuario Anni LXXIII Iove paulò minor, stellisque fulgentioribus primi ordinis, aliquanto maior fulsit, quibus in Febr. & Martio equalis exitit, sic Aprili, & Maio, stellis secundae mag. successivè ita decrescendo per Iunium. In Iulio & Augusto, tertiae quantitatis parvuit. In Octob. & Nov. undecimae Cassiepeae stellae non dispar cernebatur. In fine anni & Ianuario seq. stellas quintae formae vix excedebat. In Febr.sextas, & minimas adumbrabat, donec ultimo mense Martio, adeò exilis reddita sit, ut conspici ulteriùs prorsus defineret; nec colorem insitum eundem semper exhibuit. Tycho Brahe progym. fol. 301. Latitudinem habuit P. 53. S. 55. aliis 45. Longitudinem P.36. S. 54. dodecatemory Tauri. Stabat limbo lacteae viae Boreo affixa, ut cum stellis tertiae formae. Alfa.beta.gamma. rhombum effigeret ferè perfectum.*

*Cum nullam habuerit paralaxin, secundum proprium locum, in principio, id est Novembri, maior fuit terra. 361 ½. Sole verò 2 2/3 cuius aplituto, terrae globum, non amplius quam 139. Ut summum 140. superat & hoc ex probabiliori recentiorum sententia.* Let us note that, according to Bayer, the star was visible from November 9 (for the dates of the Julian Calendar, see please http://penelope.uchicago.edu/~grout/encyclopaedia_romana/calendar/juliancalendar.html).

Thanks to the United States Naval Observatory (USNO), we can also see the Johannes Bayer's Uranometria of 1661. It is told that the engraver was Alexander Mair, b. ca. 1559.

Actually, what we find in Bayer's Atlas of Stars is the description that was given in the Brahe's work published in 1602 by Kepler [19], as explained by Seth Ward (1617-1689), English mathematician, astronomer, and bishop [21]. In his book, Ward is also mentioning the Kepler's nova and other stars too. *Ista quidem Hipparchi stellae nouae observatio diu multumque in quaestionem a multis vocata est, donec Anno 1572. praeclara ista in Cassiiepea peregrina, novorum istiusmodi dubitationem omnem sustulerit. Cujus Historiam a Tychonis progymnasmatis in Ouranometriam suam (praeclarum opus) Bayerus ita transcripsit. Primum (inquit) a.d. 5 Id. Novemb. animadversa, veneris stellam quantitate visibili aequabat, per Decembrem Jovis Sydus ferme aemulabatur, Januario Anni 73. ... Stabat limbo lacteae viae Boreo affixa neque unquam vel longitudinem vel latitudinem mutavit Haec inquam Cassiepeae stella clarissima, quae terram ipsam Magnitudine plusquam sexcenties excessit Hipparcho fidem prima asseruit; ex quo tempore etiam novae aliae in pectore Cygni 1600. in Serpentario 1604. a Keplero sunt observatae.*

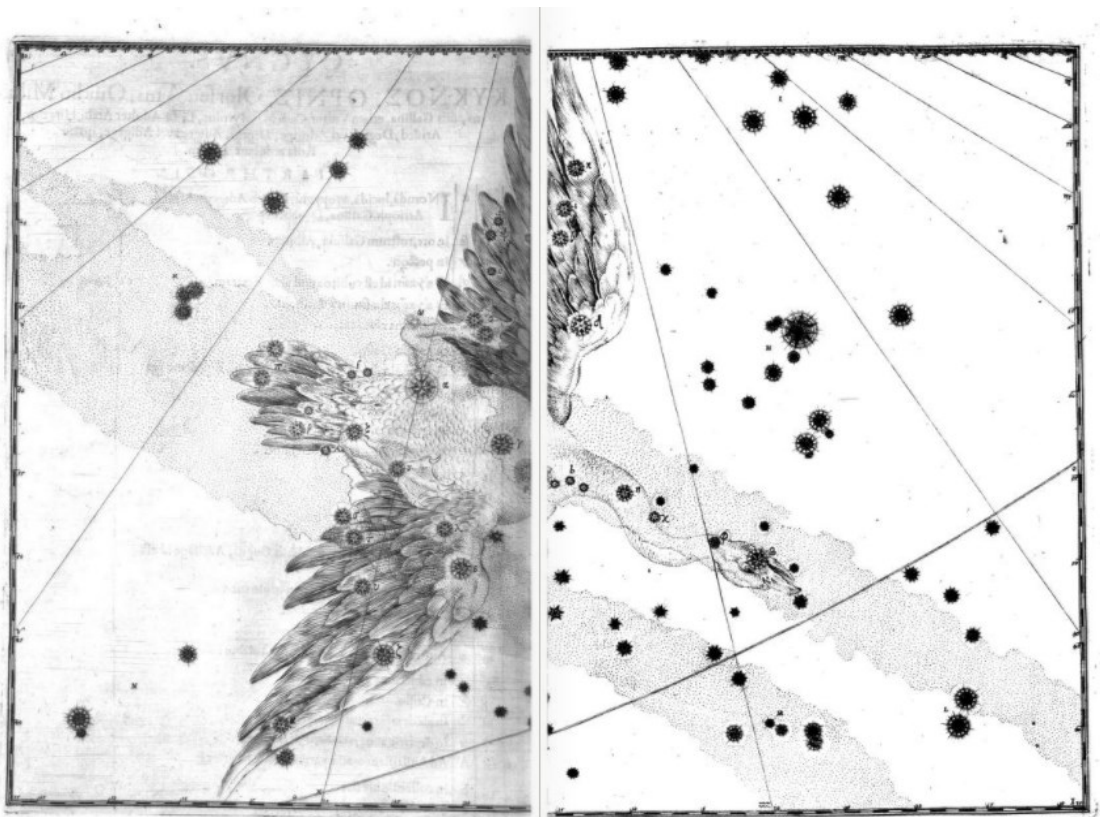

Figure 10: Constellation Cygnus in Bayer's Uranometria.

**P Cygni and other variable stars**

The star in the constellation Cygnus mentioned by Ward is P Cygni (34 Cyg), a variable star. The designation "P" was originally assigned by Bayer in his Uranometria as a new phenomenon: *Tertii fulgoris stella, Anno MDC. Primum conspecta et observata, ominium fermè tacito consensus per novo phaenomeno recepta* [20] (see also the Figures 10 and 11). As told in [22], it is a hypergiant luminous blue variable (LBV) star of spectral type B1Ia+ that is one of the most luminous stars in the Milky Way.

The star "was unknown until the end of the 16th century, when it suddenly brightened to 3rd magnitude" [22], as we can read in the Uranometria. "It was first observed on 18 August (Gregorian) 1600 by Willem Janszoon Blaeu, a Dutch astronomer, mathematician and globe-maker. Bayer's atlas of 1603 assigned it the miscellaneous label P and the name has stuck ever since. After six years the star faded slowly, dropping below naked-eye visibility in 1626. It brightened again in 1655, but had faded by 1662. Another outburst took place in 1665; this was followed by numerous fluctuations. Since 1715 P Cygni has been a fifth magnitude star, with only minor fluctuations in brightness. Today it has a magnitude of 4.8, irregularly variable … P Cygni has been called a "permanent nova" because of spectral similarities and the obvious outflow of material, and was once treated with novae as an eruptive variable; however its behaviour is no longer thought to involve the same processes associated with true novae". [22]

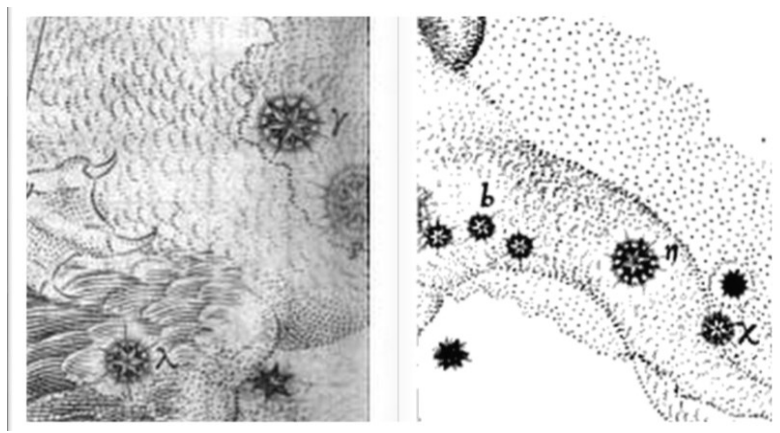

Figure 11: P Cygni in Uranometria.

In the book written by Ward [21], we read also that *Justus Pyrgius Lantg. Hass. automatopaeus in Antinoo; Simon Marius in Cingulo Andromedae; David Fabricius 1596. in Ceto sat procul a Galaxia, stellam novam observarunt. Verum in istarum stellarum Symptomatis enarrandis, naturis investigandis (quandoquidem illae cum nostro phaenomeno praeter lumen ipsum vix quicquam commune habuerint) non est ut more/mur.*

Let us consider the other stars mentioned by Ward. We read in [23] also *Non est etiam perpetuum, novas fixas Lactae adseri. Simon Marius Anno 1612, unam in cingulo Andromeda, Justus Pyrgius Landgravii Hassia: Automatopaeus alteram in Antinoo, Keplerus Ao. 1602. aliam in pisce, David Fabricius anno 1596 in Ceto, sat procul a Galaxia, nisi caligo aut lippitudines eos fefellerint, sunt conspicati.*

Justus Pyrgius was Jobst Bürgi (1552 –1632), active primarily at the courts in Kassel and Prague, a Swiss clockmaker - that is Automatopaeus - and maker of astronomical instruments [24]. In [25], we find that he was "collaboratore dal 1579 del langravio Guglielmo IV di Assia-Cassel all'osservatorio di Cassel (il primo dotato di una cupola girevole)," and he "costruì un orologio per determinare con esattezza il tempo delle osservazioni celesti. Dal 1592 lavorò a Praga per Rodolfo II. Pubblicò (1620) tavole logaritmiche".

In 1612, had Jobst Bürgi seen a supernova in Antinous? Antinous is an obsolete constellation no longer in use by astronomers. Today the asterism is merged into Aquila [26,27]. The constellation was created by Hadrian in 132, in memory of the beautiful youth, Antinous, loved by the emperor. In [27], it is told that two major novae have been observed in Aquila: the first in 389 BC and the other (Nova Aquilae 1918). A supernova in 1612 is not mentioned. Another possibility is that Jobst Bürgi has seen η Aql. It is among "the brightest Cepheid variable stars, it has a minimum magnitude of 4.4 and a maximum magnitude of 3.5 with a period of 7.2 days". Actually, in 1596 David Fabricius (1564 – 1617), a German pastor, discovered the first non-supernova variable star, Omicron Ceti, also known as Mira. At first, he believed it to be another nova [28]. "When he saw Mira brighten again in 1609, however, it became clear that a new kind of object had been discovered in the sky" [28].

Ward in [21] is also mentioning the object observed in the *cingulo Andromedae* observed by Simon Marius. Marius was the Latinized name of Simon Mayr (1573 – 1625), a German astronomer. In 1614, Marius published his work Mundus Iovialis describing the planet Jupiter and its moons [29]. "Here he claimed to have discovered the planet's four major moons some days before Galileo Galilei. This led to a dispute with Galileo, who in Il Saggiatore in 1623 accused Marius of plagiarism" [29]. A jury in The Netherlands examined in 1903 evidences extensively and ruled in favor of Marius's independent discoveries. Let us note that the mythological names that we use for the satellites today are those given them by Marius.

In cingulo Andromedae, Marius observed by means of a telescope, the Andromeda nebula [30,31]. As told in [29], the astronomer observed the location of Tycho Brahe's supernova of 1572 and found a star there, which he estimated to be "somewhat dimmer than Jupiter's third moon".

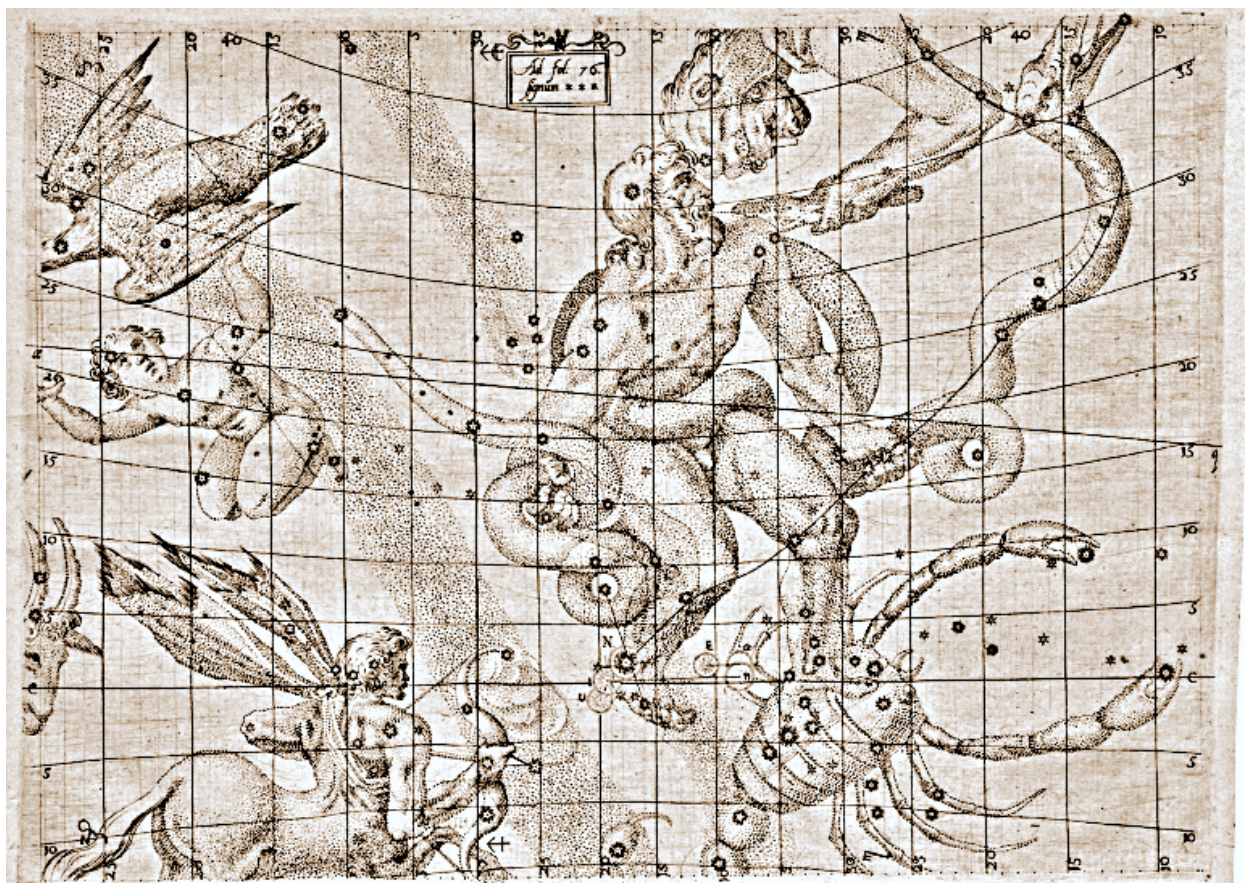

Figure 12: SN 1604 is N (that is Nova) in Ophiuchus' right foot [32].

**SN 1604**

Of course, supernova SN 1604 is not present in the Uranometria of 1603. However, we can admire it in an illustration given by Johannes Kepler [32] (see Figure 12). Today, the supernova is known as Kepler's nova. It occurred in the Milky Way, in the constellation Ophiuchus.

Appearing in 1604, "the star was brighter at its peak than any other star in the night sky, with an apparent magnitude of −2.5" [33]. Moreover, it was visible during the day for over three weeks. In [33], it is told that the first recorded observation was given by Lodovico delle Colombe in northern Italy [34]. Lodovico tells that many problems exist on which the human minds exercise themselves. *E tale appunto mi si rappresenta la materia di quelle nuove stelle, che nel Cielo si dicono essere apparite di cui fanno menzione gli Astrologi esserne state veduta una trentatrè anni sono nella sede della Cassiopea, acciò ch'io taccia la più lontana, come fu quella, che vide Hiparco; e l'Ottobre passato 1604, alli 12. ò quivi intorno un'altra nel Sagittario s'è fatta a gli occhi nostri vedere. Quella per lo corso di due anni, e questa di uno incirca si è mantenuta.*
Here again, we find mentioned Hipparchus' nova: "Hipparchus, about 135 years before the Christian era, saw a new star in the heavens (Pliny, I. ii. c. 24), which is the first of this description that has been recorded" [35]. Johannes Kepler began observing SN 1604 on October 17 when he was at the imperial court in Prague for Emperor Rudolf II [33].
This supernova again produced a series of manuscripts.

**Searching in Google Books again**
Here in the following some references.

1605 - Discorso sopra la stella nuova comparsa l'ottobre prossimo passato, dell'Eccellentissimo Astrologo e Medico Astolfo Arnerio Marchiano. In Padoa. *Varie e molto differenti s'hanno udito, & s'odono tutta via l'opinioni degli huomeni intorno alla Stella novamente l'Ottobre prossimo passato comparsa. ... tra i dotti, & valenti professori delle scienze mondane è un tumulto implacabile, errore inestricabile; & tra loro questa Stella ha causato già leggere discordie ... Stanno saldi i Filosofi volendo pur sostenere il parere d'Aristotele loro maggiore, dicendo, che nei Cieli non possono generarsi Stelle nuove, né altre maraviglie; & perciò affermano, che questa nostra non è Stella, ma una porzione d'aria accesa non molto lontana dalla terra: & non si ricordano questi; ch'al tempo della Natività del Nostro salvatore apparve una stella nuova osservata da quelli tre non manco dotti ...*

1605 - Dialogo de Cecco di Ronchitti da Bruzene in perpuosito de la stella Nuova. Al Lostrio e Rebelendo Segnor Antuogno Squerengo degnetissemo Calonego de Paua, so paròn. The Dialogue of Cecco di Ronchitti of Brugine concerning the New star is an early 17th-century pseudonymous pamphlet ridiculing the views of some Aristotelian philosophers on the nature and properties of Kepler's Supernova [36]. "It was written in the coarse language of a rustic Paduan dialect … an English translation by Stillman Drake was published in 1976. Scholars agree that the pamphlet was written either by Galileo Galilei or one of his followers, Girolamo Spinelli, or by both in collaboration" [36].

1605 - De noua stella disputatio Io. Heckij I. Lyncaei Dauentriensis philosophiae, & medicinae doctoris. Ad illustriss. dominum D. Federicum Caesium marchionem Monticellorum, &c. Joannes Heckius apud Aloisium Zannettum. Johannes van Heeck, (1579 - c.1620), also known as Johann Heck, Joannes Eck, Johannes Heckius, Johannes Eckius and Giovanni Ecchio, was a Dutch physician, naturalist, alchemist and astrologer. "Together with Prince Federico Cesi, Anastasio de Filiis and Francesco Stelluti, he was one of the four founding members of the Accademia dei Lincei" [37]. *Circa nonum Octobris diem coepit novum, ac magnum e coelo conspici portentum, lumen clarum, syderis instar, ut sydus etiam Doctorum vocetur testimonijs, fulgentibus, undequaque radijs, loco & sito Sagitarij, quemadmosum longa linearum & arcuum ferie pro Thiconico calculo fatis praeter lapso tempore designavi, ...*

1605 - Discorso di Raffael' Gvalterotti ... sopra l'apparizione de la nvova stella: E sopra le tre oscurazioni del sole, e de la luna nel anno 1605. Raffaello Gualterotti, Cosimo Giunti editore.

1605 - Iudicium oder Bedencken vom neuen Stern, welcher den 2. Oktober erschinen. Helisaeus Röslin.

1606 - De sydere novo, seu de nova stella, quae ab 8. die Oct. anni 1604 inter astra Sagittarii videri coepit: enarratio apodeictica. Autore Elia Molerio, Theologi, atque Astronomo. Excudebat Iacobus Stoer. *Cum Anno 1604 preterito & mense Octob. ab Occasu Solis post 6. diei 15. horam sereno tempore coelum intuerer …*

1609 - Historischer, Politischer und Astronomischer natürlicher Discurs von heutiger Zeit Beschaffenheit, Wesen und Standt der Christenheit, und wie es ins künfftig in derselben ergehen werde. Helisaeus Röslin.

About the texts written by Röslin, let us consider again Ref.11. "Roeslin had known Johannes Kepler since their student days and was one of his correspondents. Roeslin placed more emphasis on astrological predictions than did Kepler, and though he respected Kepler as a mathematician, he rejected some of Kepler's cosmological principles, including Copernican theory. Kepler criticized Roeslin's predictions in his book De stella nova, on the comet of 1604, and the two kept up their arguments in a series of pamphlets written as dialogues".

**Galileo and SN 1604**
Let us conclude our discussion on the literature, by mentioning the ideas of Galileo Galilei about supernovae. As we can see from the literature, the appearance of the nova in 1604 led to disputes about the Aristotelian belief of the unalterable nature of the heaven of the stars. Galileo took an active part in the dispute. As told in [38], he delivered three lectures at Padua and prepared to publish an astronomical work. However, he did not do so, "and only a short fragment of the manuscript survives". After these lectures, Galileo had a first conflict with Baldassare Capra, who was studying astronomy at the University of Padua under the guidance of Simon Marius (Mayr had returned to Germany in 1605 [39]). "In a pamphlet printed in 1605, Capra rejected the hypothesis expounded by Galileo in three lectures on the nature of the star delivered at the University of Padua" [40]. Galileo had a second conflict with Capra, due to an attempted plagiarism of the Galilean geometric and military compass [40]. In the following year, Galileo published a full account of the case in the Difesa di Galileo Galilei contro alle Calunnie ed Imposture di Baldassar Capra [41]. We do not know the arguments of the lessons about the supernova, but from the arguments against him given by Capra, we know that Galileo had interpreted the phenomenon as an evidence of the mutability of the heavens. The new star had not any change in parallax, meaning that it was beyond the orbit of the moon.
As in the case of Tycho's Nova, the "new star" forced the revision of the ancient models of heavens and the need for having better astrometric results and star catalogues was reinforced. In turn, this new Uranometria required different and more powerful astronomical observing instruments. After five years from the supernova, in 1609, Galileo was able of building a telescope, and made his first observations of celestial objects by means of it.